\documentstyle[12pt,epsfig,rotating]{article}
\newcommand{\bce}{\begin{center}}
\newcommand{\ece}{\end{center}}
\newcommand{\beq}{\begin{equation}}
\newcommand{\eeq}{\end{equation}}
\newcommand{\bea}{\vspace{0.25cm}\begin{eqnarray}}
\newcommand{\eea}{\end{eqnarray}}

\newcommand{\ba}{\begin{array}}
\newcommand{\ea}{\end{array}}

\newcommand{\doublespace}{
    \renewcommand{\baselinestretch}{1.6}\large\normalsize}

\def\lsim{\mathrel{\rlap{\lower4pt\hbox{\hskip1pt$\sim$}}
    \raise1pt\hbox{$<$}}}         
\def\gsim{\mathrel{\rlap{\lower4pt\hbox{\hskip1pt$\sim$}}
    \raise1pt\hbox{$>$}}}         

\def\Pom{{\bf I\!P}}

\def\lsim{\mathrel{\rlap{\lower4pt\hbox{\hskip1pt$\sim$}}
    \raise1pt\hbox{$<$}}}         
\def\gsim{\mathrel{\rlap{\lower4pt\hbox{\hskip1pt$\sim$}}
    \raise1pt\hbox{$>$}}}         

\def\Pom{{\bf I\!P}}

  \advance\oddsidemargin  by -1in
  \advance\evensidemargin by -1in
  \advance\topmargin      by -0.5in
\def\lsim{\mathrel{\rlap{\lower4pt\hbox{\hskip1pt$\sim$}}
    \raise1pt\hbox{$<$}}}         
\def\gsim{\mathrel{\rlap{\lower4pt\hbox{\hskip1pt$\sim$}}
    \raise1pt\hbox{$>$}}}         

\def\Pom{{\bf I\!P}}
\def\beq{\begin{equation}}
\def\endeq{\end{equation}}
\def\arr{\begin{eqnarray}}
\def\endarr{\end{eqnarray}}
\makeindex

\doublespace
\begin{document}


\phantom{.}{ \Large \hspace{10.0cm} KFA-IKP(Th)-1996-1 \\
\phantom{.} \hspace{10.0cm} DFTT 5/96 \\
\phantom{.}\hspace{11.9cm}January
 1996\vspace{0.4cm}\\ }

\begin{center}
{\bf\sl \huge Factorization properties of the diffraction
dissociation of longitudinal photons}
\vspace{0.4cm}\\
{\bf M.~Genovese$^{a,b}$,
N.N.~Nikolaev$^{c,d}$  and B.G.~Zakharov$^{d}$
\bigskip\\}
{\it $^a$       Institut de Physique Nucl\'eaire de Lyon, Universit\'e
Claude Bernard \\
43 boulevard du 11 novembre 1918, F-69622 Villeurbanne Cedex, France
\medskip\\
$^b$ Dipartimento di Fisica Teorica, Universit\`a di Torino,\\
Via P.Giuria 1, I-10125 Torino, Italy
\medskip\\
$^{c}$IKP(Theorie), KFA J{\"u}lich, 5170 J{\"u}lich, Germany
\medskip\\
$^{d}$L. D. Landau Institute for Theoretical Physics, GSP-1,
117940, \\
ul. Kosygina 2, Moscow 117334, Russia
\vspace{1.0cm}\\ }
{\Large
Abstract}\\
\end{center}
We develop the pQCD description of the diffraction dissociation
(DD) of longitudinal photons. We demonstrate that the longitudinal
diffractive structure function does not factor into the flux of
pomerons and the partonic structure function of the pomeron, thus
defying the usually assumed Regge factorization. In contrast to DD
of the transverse photons, DD of the longitudinal photons is strongly
peaked at $\beta =1$. We comment on
duality properties of DD in deep inelastic scattering.
 \bigskip\\

\begin{center}
E-mail: kph154@zam001.zam.kfa-juelich.de
\end{center}

\pagebreak

The longitudinal structure function $F_{L}(x,Q^{2})$ is a fundamental
quantity in deep inelastic scattering (DIS). In fixed target experiments,
the measurement of $F_{L}(x,Q^{2})$ requires the comparison of DIS
at different energies of the lepton, at HERA one needs to vary
either the lepton or proton beam energy or both. The exceptional
case is diffraction dissociation (DD) of photons, which can be viewed as
DIS on pomerons radiated by protons. Here the longitudinal diffractive
structure function $F_{L}^{D}(x_{\Pom},\beta,Q^{2})$ could readily be
measured varying the energy and/or $x_{\Pom}$ of the target pomeron at
a fixed energy of the electron and proton beams. The experimental
measurement of $F_{L}^{D}(x_{\Pom},\beta,Q^{2})$ could shed much light
on the microscopic QCD structure of the pomeron; the corresponding
experimental data from HERA will become available soon and thus the pQCD
evaluation of $F_{L}^{D}(x_{\Pom},\beta,Q^{2})$ is one of the topical
issues in the theory of the QCD pomeron.

The subject of the present communication is  the derivation of the pQCD
relationship between the mass spectrum in DD of longitudinal photons and
the gluon structure function of the
proton. We derive the relevant pQCD
factorization scale and establish the pattern of breaking of the Regge
factorization. Our results do clearly demonstrate that treating the
pomeron as a hadronic state endowed with a well-defined flux in the
proton and a partonic structure function is illegitimate.

We discuss the diffraction dissociation (DD) of (virtual) photons
$\gamma^{*}+p \rightarrow X +p'$ into states $X$ of mass $M$ (large
rapidity gap (LRG) events) and calculate the diffractive structure
function
defined by
\arr
\left.(M^{2}+Q^{2})
{ d\sigma^{D} (\gamma^{*}\rightarrow X)
\over dt\,d M^{2} }\right|_{t=0} =
{  \sigma_{tot}(pp) \over 16\pi}
{4\pi^{2} \alpha_{em}
\over Q^{2}}\left\{
 F_{T}^{D}(x_{\Pom},\beta,Q^{2})
+\varepsilon_{L} F_{L}^{D}(x_{\Pom},\beta,Q^{2})\right\}\, .
\label{eq:1}
\endarr
Here $Q^{2}$ is the virtuality of the photon, $W$ and $M$
are c.m.s. energy in the photon-proton and photon-pomeron
collision, $\beta =Q^{2}/(Q^{2}+M^{2})$ is the Bjorken
variable for the lepton-pomeron DIS,
$x_{\Pom}=(Q^{2}+M^{2})/(Q^{2}+W^{2})=x/\beta$ is
interpreted as the fraction of the momentum of the proton
carried away by the pomeron, $\varepsilon_{L}$ is the longitudinal
polarization of the photon and $\alpha_{em}$ is the fine structure
constant. The dimensional normalization factor $\sigma_{tot}(pp)=40$\,mb
follows from the standard Regge theory convention \cite{NZ92,NZ94}.

An assumption which is often made is that
the pomeron can be treated as a hadronic state and
$F_{T,L}^{D}(x_{\Pom},\beta,Q^{2})$ can be factored
into the partonic structure function of the pomeron
$F_{2\Pom}(\beta,Q^{2})$ and the flux of pomerons
$\phi_{\Pom}(x_{\Pom})/x_{\Pom}$ in the proton \cite{Ingelman,Regge}:
\arr
F_{T,L}^{D}(x_{\Pom},\beta,Q^{2})=
\phi_{\Pom}(x_{\Pom})
 F_{T,L}^{\Pom}(\beta,Q^{2})\, .
\label{eq:2}
\endarr
This Ingelman-Schlein-Regge factorization, which has never been derived
from the QCD analysis, involves a set of very strong assumptions
on the diffractive cross section $d\sigma^{D}$:
i) the $x_{\Pom}$ dependence is reabsorbed entirely in the
$Q^{2}, \beta$ and flavour independent pomeron flux
function $\phi_{\Pom}(x_{\Pom})$, ii) the $\beta, Q^{2}$ and
flavour dependence of $d\sigma_{D}$ are contained entirely
in the structure function of the pomeron, iii) the ratio
$R^{D}=F_{L}^{D}(x_{\Pom},\beta,Q^{2})/
F_{T}^{D}(x_{\Pom},\beta,Q^{2})$ does not depend on $x_{\Pom}$.
The purpose of the present communication is the pQCD derivation
of $F_{T,L}^{D}(x_{\Pom},\beta,Q^{2})$ and the
demonstration that none of the above properties i) to iii) holds
in the pQCD.

Different aspects of the non-factorization in DD have already
been discussed in \cite{NZ92,NZ94,NZsplit,GNZ95,GNZcharm}; the
non-factorizable colour dipole approach to DD \cite{NZ92,GNZ95}
is well known to provide a very good quantitative description of the
HERA data on LRG events \cite{H1F2Pom,ZEUSF2Pom}.

We start with diffraction excitation of photons into $q\bar{q}$ pairs,
which dominates at large $\beta$ and can be associated with DIS on the
"valence"
$q\bar{q}$ component of the photon. The formalism necessary
for our purposes has been set up in \cite{NZ92,NZsplit,GNZcharm}.
The relevant pQCD diagrams for the colour singlet exchange in the
$t$-channel are shown in Fig.~1. The mass of the
diffractively excited state $X$ is given by
$
M^{2}=(m_{f}^{2}+k^{2})/z(1-z)\, ,
$
where $m_{f}$ is the quark mass, $\vec{k}$ is the transverse
momentum of the quark with respect to the $\gamma^{*}$-pomeron
collision axis  and $z$ is the fraction of light--cone momentum
of the photon carried by the (anti)quark. Other useful kinematical
variables are
$
\varepsilon^{2}=z(1-z)Q^{2}+m_{f}^{2}$ and
\beq
q^2=
k^{2}+\varepsilon^{2}=(k^{2}+m_{f}^{2}){M^{2}+Q^{2}\over M^{2}} \,
\label{eq:3}
\endeq
After the standard leading log$\kappa^{2}$ resummation, the cross
sections of the forward ($t=0$) DD of longitudinal photons
takes the compact form \cite{NZ92,NZsplit}
\arr
\left.{d\sigma_{L} \over
dM^{2}dk^{2}dt}\right|_{t=0}={\pi^{2}\over 6}
e_{f}^{2} \alpha_{em}
\alpha_{S}^{2}(q^{2})
\cdot {Q^{2}(m_{f}^{2}+k^{2})^{3} \over
M^{7}\cos\theta \sqrt{M^{2}-4m_{f}^{2}}}
\Phi_{2}^{2}  \, .
\label{eq:4}
\endarr
Here $e_{f}$ is the quark charge in units of the electron change,
$\theta$ is the quark production angle with respect to the
$\gamma^{*}$-pomeron collision axis,
\beq
\Phi_{2}=\int {d\kappa^{2}\over \kappa^{4}}
f(x_{\Pom},\kappa^{2})
\left[{1\over \sqrt{a^{2}-b^{2}}} -
{1\over k^{2}+\varepsilon^{2}}
\right] \, ,
\label{eq:5}
\endeq
$a=\varepsilon^{2}+k^{2}+\kappa^{2}$,
$b=2k\kappa$
and $f(x_{\Pom},\kappa^{2})=\partial
G(x_{\Pom},\kappa^{2})/\partial \log \kappa^{2}$
is the unintegrated gluon structure function of the target proton.
Following the analysis \cite{NZsplit,GNZcharm} one can easily verify
that after factoring out $(k^{2}-\epsilon^{2})/(k^{2}+m_{f}^{2})^{3}$
in (\ref{eq:5}), one will be left with the logarithmic
$\kappa^{2}$ integration with $q^{2}$ being the upper limit
of integration. Consequently, $q^{2}$ emerges as the pQCD factorization
scale (it has already been used as such in the running strong coupling
$\alpha_{S}(q^{2})$ in (\ref{eq:4})) and to the leading log$q^{2}$,
\beq
\Phi_{2}=
{M^{4}[(k^{2}+m_{f}^{2})(M^{2}-Q^{2})-2m_{f}^{2}M^{2}]
\over
(Q^{2}+M^{2})^{3}(k^{2}+m_{f}^{2})^{3} }
G(x_{\Pom},q^2)\, .
\label{eq:6}
\endeq
Notice a zero of the $d\sigma_{L}$ at
$
(k^{2}+m_{f}^{2})(M^{2}-Q^{2})=2m_{f}^{2}M^{2}$.
For light flavours, $d\sigma_{L}$ vanishes at $M^{2}=Q^{2}$.
Substituting (\ref{eq:6}) into (\ref{eq:5}), one readily
finds
\arr
\left.{d\sigma_{L} \over
dM^{2}dk^{2}dt}\right|_{t=0}={\pi^{2}\over 6}
e_{f}^{2} \alpha_{em}
\alpha_{S}^{2}(q^{2})
G^{2}(x,q^{2}) {Q^{2}M[(k^{2}+m_{f}^2)(M^{2}-Q^{2})-2m_{f}^{2}M^{2}]^{2}
\over
\cos\theta (Q^{2}+M^{2})^{6}(k^{2}+m_{f}^{2})^{3}
\sqrt{M^{2}-4m_{f}^{2}} }\, .
\label{eq:7}
\endarr
Notice that in the DIS limit of $Q^{2} \gg m_{f}^{2}$, the $k^{2}$ and
$M^{2}$ dependences in (\ref{eq:7}) do factor, which leads to the simple
$\beta$ dependence  $F_{L}^{D} \propto (1- 2 \beta)^2 \beta ^3$. The
r.h.s. of Eq.~(\ref{eq:7}) decreases with $k^{2}$ only as $k^{-2}$ and
one has the logarithmic integration $\int^{{1\over 4}M^{2}}dk^{2}/(k^{2}
+m_{f}^{2})$. The Jacobian peak singularity and the scaling violations
in $G(x,q^{2})$ further enhance the contribution from
large $k^{2} \sim {1\over 4}M^{2}-m_f^{2}$. Consequently,
the relevant pQCD factorization scale
equals
\beq
q^{2}\approx {1\over 4\beta} Q^2 .
\label{eq:8}
\endeq
The $k^{2}$ integration produces a logarithmic factor of the
form $\log(M^{2}/4m_{f}^{2}) = \log(Q^{2}(1-\beta)/4\beta m_{f}^{2})$.
This factor has only a marginal effect on the $\beta$ and $Q^{2}$
dependence. At asymptotically large $Q^{2}\gg 4m_{f}^{2}$ the
flavour symmetry is restored, but for $Q^{2}$ of practical
interest there is a substantial suppression of the charm cross section,
similar to the suppression of the charm structure function of the
proton \cite{F2charm}.
Suppressing this factor, to a logarithmic accuracy,
\beq
F_{L}^{D}(x_{\Pom},\beta,Q^{2})=e_{f}^{2} {2 \pi \over 3\sigma_{tot}(pp)
Q^{2}}
(1- 2 \beta)^2 \beta ^3 \cdot
\alpha_{S}^{2}({Q^{2}\over 4\beta}) \cdot G^{2}({Q^2 \over 4
\beta},x_{\Pom})^2\, ,
\label{eq:9}
\endeq
which concludes the derivation of the longitudinal diffractive
structure function. Because of the zero at $\beta={1\over 2}$ and
the $\beta^{3}$ dependence, the $F_{T,L}^{D}(x_{\Pom},\beta,Q^{2})$
is strongly peaked at $\beta=1$, so that one can put, with a good
accuracy, $\beta=1$ in the factorization scale.

The salient features of the $F_{L}^{D}(x_{\Pom},\beta,Q^{2})$ are clearly
seen from Eq.~(\ref{eq:9}). First, it is short distance dominated and is
exactly calculable in the realm of pQCD \cite{NZ91,NZ92}. Second, it has
the higher twist dependence $\propto 1/Q^{2}$, a result known since
\cite{NZ91}. Third, the $\beta$ and $x_{\Pom}$ dependences do factorize.
Fourth, the $x_{\Pom}$- and $Q^2$-dependences are inextricably
entangled, the Regge factorization (\ref{eq:2}) breaks down and
neither the concept of a $Q^{2}$ independent
flux of pomerons nor the one  of a pomeron structure function (which absorbs
all the $Q^{2}$ dependence) do make sense.
Regretfully, these concepts
have become customary in the analysis and presentation of the experimental
data. The above pQCD derivation shows unequivocally that if one wants to
keep the pomeron structure function language, then one can do so only
at the expense of modifying the Eq.~\ref{eq:2} to allow for the $Q^2$
dependent pomeron flux function:
\beq
\phi^L_{\Pom}(x_{\Pom},Q^2) =\left ( { G(x_{\Pom}, Q^2/4) \over
G(x_{0}, Q^2/4)} \right )^2\, .
\label{eq:10}
\endeq
Here the normalization is $\phi^L_{\Pom}(x_{0}=0.03)=1$
for every $Q^2$
\cite{GNZ95}.
Notice that the so defined $\phi^L_{\Pom}(x_{0},Q^2)$
is flavour independent, in contrast to the diffraction dissociation
of transverse photons where excitation of each and every new flavour
entails the brand new pomeron flux function \cite{GNZcharm}. Then, with
all the above reservations, one can define the longitudinal structure
function of the pomeron,
\beq
 F_{2\Pom}^{L}(\beta)= \Sigma_{f} e_{f}^{2}{A_{f} \over Q^2} (1-2 \beta)^2
 \beta^3\, ,
\label{eq:11}
\endeq
where the normalization factors $A_f$ are, to a first approximation,
$Q^2$ independent.

For the numerical evaluation of the longitudinal cross section
and the normalization factors $A_f$ in (\ref{eq:11})
we rely upon the colour dipole gBFKL formalism \cite{NZ91,NZ92,NZ94}.
The $M^{2}$ and/or $\beta$ integrated DD cross section equals
\beq
\left. { d \sigma_{L}^{D} \over dt } \right|_{t=0} =
\int dM^{2}\left. { d \sigma_{L}^{D} \over dt dM^{2} } \right|_{t=0} =
 {1\over 16 \pi}\int_{0}^{1}dz \int d^{2}\vec{\rho}\,
\vert\Psi_{\gamma^{*}}^L(Q^{2},z,r)\vert^{2} \sigma^{2}(x,r)
\label{eq:12}
\endeq
where
\beq
\vert\Psi_{\gamma^{*}}^L(Q^{2},z,r)\vert^{2}
={6\alpha_{em} \over (2\pi)^{2}}
\sum_{i}^{N_{f}}e_{f}^{2}
4 Q^2 z^2 (1-z)^2
K_{0}(\varepsilon r)^{2}
\label{eq:13}
\endeq
gives the colour dipole distribution in longitudinal photons
\cite{NZ91} and $\sigma(x,r)$ is the colour dipole cross section
from ref. \cite{NZHera,NNZscan}. The resulting
$\phi^{L}_{\Pom}(x_{\Pom},Q^{2})$
can conveniently be parameterized
(for $1 \lsim Q^2 \lsim 100 GeV^2$) as
\beq
\phi^{L}_{\Pom}(x_{\Pom},Q^{2})= \left({x_0 \over x_{\Pom}} \right )
^{\left[a+d \log (Q^2/10)+ f \log^2 (Q^2/10) \right ]}
\cdot \left [{ x_{\Pom} + c \over x_0 + c}\right ]
^{\left [b + e \log (Q^2/10)\right]}
\label{eq:14}
\endeq
where $Q^{2}$ is in GeV$^{2}$, $a=0.456$, $b=0.678$, $c= 0.012$,
$d=0.112$, $e=0.078$ and $f=0.01$.

The generalized flux function $\phi^{L}_{\Pom}(x_{\Pom},Q^{2})$
is flavour independent, the flavour dependent normalizations
$A_{f}$ in (\ref{eq:11}) can be determined
equating the cross sections given
by Eq.~(\ref{eq:12}) and the $M^{2}$ and/or $\beta$-integrated
Eq.~(\ref{eq:9}). In the interesting range of $Q^{2} \lsim 100$
GeV$^{2}$, the result is
: $A_{ud}=0.82$,
$A_{s}=0.61$ and $A_{c}=0.05$. Notice that $A_{ud} \approx A_{s}$,
for the charm the flavour symmetry is strongly broken. This order
of magnitude estimate for $A_{c}$ is sufficient for evaluations
of the numerically small charm cross section.

A very different situation occurs for the triple pomeron region
of $\beta \ll 1$, which is dominated by DD into  $q\bar{q}g...$
states. The factorization properties of DD in this region of
$\beta$ can clearly be seen from diffractive excitation of the
$q\bar{q}g$ states of the photon, which gives the driving term
of $F_{T,L}^{D}(x_{\Pom},\beta,Q^{2})$ at $\beta\ll 1$. In \cite{NZ94}
it has been shown
that DD in DIS is dominated by configurations in which the transverse
separation $\rho$ of the gluon from the $q\bar{q}$ pair is much
larger than the $q$-$\bar{q}$ separation $r$. Then, the DD cross
section can be factored as
\arr
(Q^{2}+M^{2})\left.{d\sigma^{D}_{T,L} \over dt dM^{2}}\right|_{t=0}
\simeq \int dz \,d^{2}\vec{r}\,\,
|\Psi_{\gamma^{*}}^{T,L}(Q^{2},z,r)|^{2}\cdot {16\pi^{2} \over 27}
\cdot\alpha_{S}(r) r^{2} \nonumber\\
\times {1\over 2\pi^{4}}\cdot\left({9\over 8}\right)^{3}
\cdot \int d\rho^{2}
\left[{\sigma(x_{\Pom},\rho)\over \rho^{2}}\right]^{2}
{\cal F}(\rho)\, ,
\label{eq:15}
\endarr
where ${\cal F}(\rho)$ provides an infrared cutoff at distances $\rho$
exceeding the propagation radius for perturbative gluons, for a
detailed discussion see \cite{NZ94,GNZ95,GNZA3Pom}. The crucial point
is that the $x_{\Pom}$ dependence in (\ref{eq:15}) decouples from
the $Q^{2}$ and $\beta$ dependence and is universal for the
$d\sigma_{T}^{D}$ and $d\sigma_{L}^{D}$
(as well as flavour independent)
as soon as $Q^{2}\gsim
3$\,GeV$^{2}$ \cite{GNZA3Pom}, it is given by the pomeron
flux function $f_{\Pom}(x_{\Pom})$  calculated in
\cite{GNZ95}. Eq.~(\ref{eq:15}) gives the
driving term of the leading-log${1\over \beta}$  expansion
of the DD cross section. It can be argued that at least to the
leading-log${1\over \beta}$, the diffractive structure function
has the conventional GLDAP evolution properties, the corresponding
analysis \cite{NZ94} needs not be repeated here. The structure of
the $r$ integrations in (\ref{eq:15}) is only marginally different
from that in the DIS structure function at small $x$. The detailed
calculation of the ratio $R^{DIS}=\sigma_{L}/\sigma_{T}$ for DIS
has been performed in \cite{NZ91,NZHera}, the major
finding is that $R^{DIS}\approx 0.2$ with a very weak $Q^{2}$ and
$x$ dependence. Consequently, we expect a close similarity of
$R^{D}$ at small $\beta$ to $R^{DIS}$ at small $x$. In Fig.~2 we
present our results for $R^{D}$ at $\beta \ll 1$
 as a function of $Q^{2}$, as it
was anticipated it exhibits very weak $Q^{2}$ dependence, with
the exception of the excitation of open charm, where the standard
threshold behaviour $\propto Q^{2}/(Q^{2}+4m_{c}^{2})$ is clearly seen.

For a numerical estimate of the longitudinal diffractive structure
function we will use in
the following: for the valence part,
Eq.~(\ref{eq:11}) with the flux parametrization Eq.~(\ref{eq:14})
(which reproduces the exact result to a $\approx 10 \%$ accuracy
in the $(x_{\Pom},Q^2)$ region relevant at HERA,
$10^{-4} < x_{\Pom} < 0.03$, $1 $ GeV$^2 \leq Q^2 \leq 100 $GeV$^2$);
for the sea component the results for $F^D_T$ from Ref.
\cite{GNZ95,GNZcharm}
assuming $R^D=0.2$ constant in the whole region.

In Fig.~3 we show how the transverse and longitudinal diffractive
structure functions $F_{T,L}^{D}(x_{\Pom},\beta,Q^{2})$ evolve with
$Q^{2}$ and $x_{\Pom}$. The $Q^{2}$ evolution of $F_{T}^{D}$ is marginal,
we show it for $Q^{2}=10$\,GeV$^{2}$. At small $\beta \ll 1$, the
longitudinal contribution is small, $R^{D}\approx 0.2$, and both the
$F_{T}^{D}$ and $F_{L}^{D}$ have identical $x_{\Pom}$ dependence.
As $F_{T}^{D}(x_{\Pom},\beta,Q^{2})$ vanishes for
$\beta \rightarrow 1$, at $\beta\gsim 0.9$ the diffractive
structure function is entirely dominated by the
$F_{L}^{D}(x_{\Pom},\beta,Q^{2})$. At fixed $x_{\Pom}$, the
longitudinal structure function decreases with $Q^{2}$, however
at small $x_{\Pom}$ the higher twist behaviour $F_{L}^{D}\propto 1/Q^{2}$
is to a large extent compensated by the scaling violations in the
gluon structure function. For this reason, $F_{L}^{D}$ remains
non-negligible even for $Q^{2}$ as large as $Q^{2} \sim 100$\,GeV$^{2}$.
Notice also the steeper $x_{\Pom}$-dependence of
$F_{L}^{D}(x_{\Pom},\beta,Q^{2})$ at large $\beta$ as compared to
the $x_{\Pom}$ dependence of $F_{T}$.
In the typical kinematics of the HERA experiments
$\epsilon_{L}\approx 1$ and the measured diffractive structure
function roughly corresponds to $F_{2}^{D}=F_{T}^{D}+F_{L}^D$. In the
range $x_{\Pom}=[10^{-3},10^{-2}]$ of the present HERA experiments,
the $x_{\Pom}$ dependence of $F_{2}^{D}$ can be parametrized by
the law $\propto x_{\Pom}^{-\delta}$ to a $\approx 20 \%$ accuracy.
The so estimated exponent $\delta$ is shown in Fig.~4 for
$Q^{2}=100$\,GeV$^{2}$ and $Q^{2}=10$\,GeV$^{2}$.
The  exponent $\delta$ rises towards $\beta \rightarrow 1$,
takes a minimal value  at moderately small $\beta$, then rises
again towards small $\beta$. The approximation
$\propto x_{\Pom}^{-\delta}$ for the $x_{\Pom}$ dependence is
rather crude; the value of the exponent $\delta$ depends on
the range of $x_{\Pom}$, the explicit form of the $x_{\Pom}$
dependence is shown in \cite{GNZ95,GNZcharm}. The values of $\delta$
evaluated for the range $x_{\Pom}=[10^{-3},3\cdot 10^{-2}]$ are
uniformely lower by $\approx $0.03-0.04 than those shown in Fig.~3,
however the form of the $\beta$ dependence of the exponent $\delta$ is
fully preserved.
Because of the partial compensation, which has been described
previously, of the higher twist
behaviour of $F_{L}$ by the scaling violations in the generalized
flux (\ref{eq:10}), (\ref{eq:14}), the rise of the exponent $\delta$
towards $\beta \rightarrow 1$ persists at all the $Q^{2}$ and is quite
relevant.

Finally, we wish to comment on the Bloom-Gilman-Drell-Yan-West
duality-type relationship between the diffraction dissociation into
the $q\bar{q}$ continuum at $\beta \rightarrow 1$ and the exclusive
diffractive production of vector mesons:
\beq
\int_{0}^{M_{V}^{2}} dM^{2}
{d \sigma_{T,L}^{D} \over dM^{2}}
\propto {1\over Q^{2}}\int_{\beta_{0}}^{1}d\beta
F_{T,L}^{D}(x,\beta,Q^{2})
\propto
\sigma(\gamma^{*}_{T,L}N\rightarrow V_{T,L}N)
\label{eq:16}
\endeq
In the l.h.s. of (\ref{eq:16}), the integration goes over the resonance
mass range $M^{2} \sim M_{V}^{2}$ and/or over the large-$\beta$ domain
$1-\beta \lsim 1-\beta_{0} \sim {M_{V}^{2} \over Q^{2}}$. In this
domain, $x_{\Pom}$ coincides with the Bjorken variable $x$. The pQCD
description of exclusive production of vector mesons has been developed
in \cite{KNNZ94}. The longitudinal photons produce longitudinally
polarized vector mesons, in \cite{KNNZ94}  it was shown
that $\sigma(\gamma^{*}_{L}N\rightarrow V_{L}N) \propto Q^{-6}
G^{2}(x,\tau Q^{2})$, the factor $\tau \sim 0.1-0.2$ in the
factorization scale was derived in \cite{NNZscan}. One recovers precisely
the same $x$ and $Q^{2}$ dependence after the integration of the mass
spectrum (\ref{eq:7}) over $M^{2}\lsim M_{V}^{2}$, both the higher twist
behaviour and the flat $\beta$ dependence at $\beta \rightarrow 1$ in
(\ref{eq:9}, \ref{eq:11}) are crucial for this consistency. Similar
consistency with the duality is found for the transverse photons.
Namely, here the result of ref.~\cite{KNNZ94,NNZscan} for the exclusive
cross section is $\sigma(\gamma^{*}_{T}N\rightarrow V_{T}N) \propto
Q^{-8}G^{2}(x,\tau Q^{2})$. The limiting behaviour of the mass spectrum
for DD of transverse photons at $\beta \rightarrow 1$ has been derived
in our previous paper \cite{GNZcharm},
\beq
F_{T}^{D}(x_{\Pom},\beta,Q^{2}) \propto (1-\beta)^{2}G^{2}(x,q^{2})\, ,
\label{eq:17}
\endeq
where the factorization scale $q^{2}$ equals
 \beq
q^{2} \sim {m_{f}^{2}\over 1-\beta }=m_{f}^{2}(1+{Q^{2}\over M^{2}})\, .
\label{eq:18}
\endeq
In the exclusive limit, $q^{2}\propto Q^{2}$ and
Eqs.~(\ref{eq:17}),(\ref{eq:18}) entail the identical $x$ and
$Q^{2}$ dependence of the l.h.s. and r.h.s. of Eq.~(\ref{eq:16}).
Notice a remarkable conspiracy of breaking of the Regge factorization
in the longitudinal and transverse cross sections and of the
$\beta$ dependence of the pQCD factorization scales (\ref{eq:8}) and
(\ref{eq:18}), which is crucial for the duality relationship between
the exclusive vector meson production and diffraction dissociation
to hold for both the longitudinal and the transverse photons.

{\bf Summary and conclusions.} The presented QCD derivation of the mass
spectrum for diffraction dissociation of longitudinal photons completes
the analysis of the breaking of Regge factorization in diffractive
deep inelastic scattering. The present results, together with those of
our previous works \cite{NZ92,GNZ95,GNZcharm} do unequivocally
demonstrate that
the Ingelman-Schlein-Regge factorization (\ref{eq:2}) is not born
out by the QCD analysis of diffraction dissociation. The predicted
breaking of the Regge factorization is strong and we look forward to
the higher statistics data from HERA. Testing our predictions for the
longitudinal diffractive structure function will be feasible in the near
future, because the diffraction dissociation of photons is the unique
process in which one can readily separate the longitudinal and transverse
structure functions without varying the electron and proton beam energies.
\medskip\\
{\bf Acknowledgments:} M. Genovese and B.G.Zakharov thank J.Speth for the
hospitality at the Institut f\"ur Kernphysik, KFA, J\"ulich.
This work was partly supported by the INTAS grant 93-239 and
the Grant N9S000 from the International Science Foundation.
M. Genovese thanks M. Giffon for useful discussions.
\pagebreak

\pagebreak
{\bf \Large Figure captions}
\begin{itemize}
\item[Fig.1]
 - One of the 16 Feyman diagrams for diffraction excitation of
the $q\bar{q}$ state of the photon.

\item[Fig.2]
 - $R^{D}$  at $\beta \ll 1$ for the light quark (solid curve), the
strange (dashed curve) and the charm (dot--dashed curve) components.

\item[Fig.3]
 - $F^D_T(x_{\Pom},\beta,Q^2)$ and $F^D_L(x_{\Pom},\beta,Q^2)$
versus $\beta$ at $x_{\Pom}=0.03$ and $x_{\Pom}=0.0003$.
$F^{D}_{T}(x_{\Pom},\beta,Q^2)$ is shown
at $Q^2 = 10$ GeV$^2$ (solid curve), while $F^D_L(x_{\Pom},\beta,Q^2)$
is reported for $Q^2 = 10$ GeV$^2$
(dot--dashed curve), $Q^2 = 50$ GeV$^2$ (dashed curve) and
$Q^2 = 100$ GeV$^2$ (dotted curve).

\item[Fig.4]
 - The exponent $\delta$ of the $x_{\Pom}$ dependence of the
observed diffractive structure function $F^{D}=F_{T}^{D}+F_{L}^{D}$
at $Q^2=10$ GeV$^2$ (dashed curve) and at $Q^2=100$ GeV$^2$
(solid curve).

\end{itemize}
\end{document}